\documentstyle[aps,prd,multicol]{revtex}
\tighten

\begin{document}

\title{Fermion masses in a model for spontaneous parity breaking}
\author{Y. A.
Coutinho\thanks{yara@if.ufrj.br},\\
J. A.  Martins Sim\~oes\thanks{simoes@if.ufrj.br} \\ Instituto
de F\'{\i}sica,\\
Universidade Federal do Rio de Janeiro, RJ, Brazil\\
and\\
C.M. Porto\\
Universidade Federal Rural do Rio de Janeiro,
RJ, Brazil}
\maketitle
\begin{abstract}
\par
In this paper we discuss a left-right symmetric model for  elementary particles and their connection with the mass spectrum of elementary fermions. The model is based on the group $SU(2)_{L}\otimes SU(2)_{R}\otimes U(1)$. New mirror fermions and a minimal set of Higgs particles that breaks this symmetry down to $U(1)_{em}$ are proposed. The model can accommodate a consistent pattern for charged and neutral fermion masses as well as neutrino oscillations. An important consequence of the model is that the connection between the left and right sectors can be done by the neutral vector gauge boson $Z$ and a new heavy $Z'$. 
 
\end{abstract}
\begin{multicols}{2}

\section{Introduction}\setcounter{equation}{0}

\par
The origin of the left-right asymmetry in weak interactions is a longstanding problem in elementary particle physics. One possible way to understand this asymmetry is to enlarge the standard model into  a left-right symmetric structure and then, by some spontaneously broken mechanism, to recover  the low energy asymmetric world. Many models were developed, based on grand unified groups \cite  {BBR}, superstring inspired models \cite{ZGB}, a connection between parity and the strong CP problem \cite{SMB}, left-right extended standard models \cite{RFO}. All these approaches imply the existence of some new intermediate physical mass scale, well bellow the unification or the Planck mass scale. The increasing experimental evidence on neutrino oscillations and non zero masses has also motivated a renewed interest on mechanisms for parity breaking.
\par
Left-right models starting from the gauge group  $SU(2)_{L}\otimes SU(2)_{R}\otimes U(1)_{B-L}$ were developed by many authors \cite{JCP} and are well known to be consistent with the standard  $SU(2)_{L}\otimes U(1)_Y$ predictions for fundamental interactions and imply new heavy gauge bosons, not yet observed. However, for the fermion mass spectrum there is no unique choice of the Higgs sector that can reproduce the observed values for both charged and neutral fermions, neither the fundamental fermionic representation is uniquely defined. 
\par
In this paper we present a model that starts by the simple gauge structure of $SU(2)_{L}\otimes SU(2)_{R}\otimes U(1)$ and investigate a minimal Higgs sector that breaks the left-right symmetry. The fermion mass spectrum is studied for several choices of the Higgs sector. This paper is organized as follows: in section II we present our main assumptions for a left-right model; in section III we review the properties of new gauge interactions; in section IV we present a number of possibilities for the Higgs sector of fermion masses; in section V we show some phenomenological consequences for testing the models here proposed and in section VI we give our conclusions.

\section{The model}\setcounter{equation}{0}

\par
 The fundamental fermionic transformation under parity is very simple in the context of QED: it is done by the transformation between the right and left components of the same fundamental fields. Weak interactions indicates new elements in the parity transformation mechanism since the right and left components have different isospin assignments and there is no deeper understanding of this asymmetry. As the parity asymmetry is clearly displayed only in the weak isospin doublet of the fundamental fermionic representation, we take as a minimal gauge sector the left-right symmetric group $SU(2)_{L}\otimes SU(2)_{R}\otimes U(1)_Y$, with generators $(T,T^*,Y)$. The $U(1)$ quantum number could be gauged as B-L. This group can be considered as a sub-group of many unification groups like the superstring inspired $E_8\otimes E_8'$ or the SUSY - $SO(10)\otimes SO(10)'$. Earlier left-right models \cite{JCP} were based on the hypothesis of a symmetric world with fundamental fermions at the same mass scale of the presently known standard quarks and leptons. Some recent models propose a duplication of the full standard model gauge sector $SU(3)\otimes SU(2)\otimes U(1)$. More recently other models \cite{BBR,RFO} were proposed with different left and right mass scales. We consider a particular choice that avoids the introduction of a second photon, which is known to present difficulties with the positronium decay rate \cite{SLG}. We will assume this starting 
point  and introduce new mirror fermions with the following assignment:

\begin{equation}
{\ell_L={\nu \choose e}_L ,\ \nu_R ,\ e_R
\quad \buildrel {\rm P} \over \longleftrightarrow \quad L_R={N \choose E}_R,\ N_L,\ E_L}
\end{equation}

\begin{displaymath}
({1\over 2}, 0, -1); (0, 0, 0); (0, 0, -2) \quad (0, {1\over 2}, -1); (0, 0, 0); (0, 0, -2)
\end{displaymath}
\bigskip
\par
The other lepton and quark families follow a similar pattern. This fundamental representation clearly is anomaly free. In this model the parity operation transforms the $SU(2)_{L}\buildrel {\rm P} \over \longleftrightarrow SU(2)_{R}$ sectors, including the vector gauge bosons. This symmetry property can be used to reduce the three group constants $g_L;g_R;g'$ to only two, with $g_L=g_R$. For the other leptonic and quark families a similar structure is proposed. The charge generator is given by $Q = T_3+T_3^*+Y/2$.
\smallskip 
\par
In order to break $SU(2)_{L}\otimes SU(2)_{R}\otimes U(1)_Y$ down to $U(1)_{em}$ we introduce two Higgs doublets that under parity are transformed as $\chi_L \leftrightarrow  \chi_R$. Their quantum numbers are:

\begin{equation}
\chi_L \qquad \buildrel {\rm P} \over \longleftrightarrow  \quad \chi_R 
\end{equation}

\begin{displaymath}
({1\over 2}, 0, 1) \qquad (0, {1\over 2}, 1)
\end{displaymath}

\noindent
and the corresponding symmetry breaking is realized through the vacuum parameters $v_L$ and $v_R$. The model includes a Higgs field $\Phi$ in the mixed representation (1/2,1/2,0). The symmetry breaking is done by the neutral components, with parameters $k$ and $k'$. Finally we consider  scalar fields in the representation (0,0,0), broken at $s_{GUT}$ or at lower scales.
In the standard model, and also in some of its extentions, Higgs doublets are responsible for the gauge boson masses as well as for the fermion masses. For the gauge bosons this hypothesis is strongly confirmed by the experimental result $\rho=1$. For the fermion masses the standard model requires adjusting by hand the Yukawa couplings in order to reproduce the observed mass spectrum. Although this procedure is consistent in the sense that all couplings satisfy $g_i < 1$, there is no direct experimental confirmation for this hypothesis. One of the main points of our work is to show that Higgs singlets in left-right models can give a consistent charged and neutral fermion mass spectrum. We are also supposing the symmetry breaking hierarchy $s_{GUT}>>v_R>>v_L,k,k'$.
\par 
\section{The gauge bosons}\setcounter{equation}{0}

\par
In order to make our presentation more complete we briefly review in this section the main properties of the gauge sector in the $SU(2)_{L}\otimes SU(2)_{R}\otimes U(1)_Y$ model.

The symmetry breaking mechanism follows the standard model procedure and, with the above hierarchy, we readily find for the charged vector bosons the masses (for $v_L>>k,k'$):
\begin{eqnarray}
M_{W_L}^2 & \simeq &{1\over 4} g_L^2 v_L^2  \nonumber\\
M_{W_R}^2 & \simeq &{1\over 4} g_R^2 v_R^2
\end{eqnarray}

\par
We call attention to the fact that the above Higgs field in the mixed representation (1/2,1/2,0) implies a mixing between $W_L$ and $W_R$. This mixing is given by $\sin\theta_M = k k'/v_R^2$. It is well known that this mixing is strongly suppressed, as required by the absence of right-handed currents \cite{JCP,SEN}. From the first equation III.1, we have the identification $v_L=v_{Fermi}$ and the charged current interactions of standard fermions must be according to $(1/2;0;Y) \rightarrow W_L$ and for the new mirror fermions $(0;1/2;Y)\rightarrow W_R$. There are alternatives to this scheme. We can also have $v_L=0$ and the Fermi scale is given by $v_{Fermi}^2 \simeq k^2+k'^2$ \cite{SEN} with a small mixing between $W_L$ and $W_R$.

\par
Before proceeding to the neutral sector it is convenient to introduce the notation,
\begin{eqnarray}
sin^2\theta_W  & \equiv & {{g^2_R g'^2}\over {g^2_R g^2_L +g^2_R g'^2 +g^2_L g'^2}}\nonumber\\
sin^2\beta & \equiv & {{g'^2}\over{g^2_R + g'^2}}
\end{eqnarray}

\par
This is a change from ($g_L$; $g_R$; $g'$) to the basis $(g_L;\sin\theta_W;\sin\beta)$. The condition $g_L=g_R$ implies $\sin\beta=\tan\theta_W$. We also introduce the ratio $\omega= v_L/v_R$, which 
is supposed to satisfy $\omega^2\ll 1$ due to the non observation of any new physical scale with energies up to 1 TeV.
\par
For the neutral vector boson sector we find, after diagonalization and expanding in powers of $\omega$,

\begin{eqnarray}
M_{\gamma} & \equiv & 0 \nonumber\\
M^2_Z & \equiv & {1\over 4}{{v_L^2 g^2_L}\over {\cos^2\theta_W}}\lbrace 1-\omega^2 \sin^4 \beta\rbrace \nonumber\\
M^2_{Z'}& \equiv & {1\over 4}{{v^2_R g^2_L}{\tan^2\theta_W \tan^2\beta}}\lbrace 1+ {\omega^2 sin^2{2\beta}\over{4 \sin^2 \theta_W}}\rbrace 
\end{eqnarray}

\par
The unification condition for the electromagnetic interaction is the same as in the standard model,
\bigskip
\begin{equation}
e= g_L \sin \theta_W, 
\end{equation}
and the $\rho$ parameter is changed to
\bigskip
\begin{equation}
\rho = {M^2_{WL}\over {M^2_Z \cos^2 \theta_W}}\lbrace 1 - \omega^2 \sin^4 \beta\rbrace
\end{equation}
\par
In order to compare the model with the experimental data, we must develop the neutral current interactions. First we recall that the gauge sector of the model has five input parameters  $g_L$; $g_R$; $g'$; $v_L$; $v_R $. The condition $g_L=g_R$ reduces them to four. We take as experimental inputs \cite{RPP} $M_W=80.4$ GeV; $\sin^2\theta_W=0.2230$ and $\alpha(M_Z)=1/128$. The W mass is much less precise than the Z mass, but we have taken this procedure since from equation III.1 the determination of the fundamental parameters is much more transparent. If we start from the Z mass there is no significative changes in our conclusions. With the values above we readily obtain $v_L=248$ GeV and $g_L=0.6635$ as in the standard model and 

\bigskip
\begin{equation}
\sin^2\beta=0.286
\end{equation}

\par
Then the only unknown in the gauge sector of the model is $v_R$ which is supposed to be higher than $v_L$.
\par
The neutral currents coupled to the massive vector bosons Z and Z' are given by:

\begin{eqnarray}
J_{\mu} & = & {g_L \over\cos\theta_W}\gamma_{\mu}\lbrace (1-\omega^2 \sin^4 \beta)T_3- \omega^2 \sin^2 \beta \cos^2 \beta T^*_3  \nonumber\\
& - & Q \sin^2 \theta_W (1- {\omega^2 \sin^4\beta\over \sin^2 \theta_W})\rbrace \nonumber\\
J'_{\mu} & = & g_L \tan\theta_W \tan\beta\gamma_{\mu}\lbrace (1 + {{\omega^2 \sin^2\beta\cos^2 
\beta} \over \sin^2 \theta_W})T_3  \nonumber\\ 
& + &{1\over \sin^2 \beta} T^*_3 -  Q(1+ \omega^2 \cos^2 \beta \sin^2 \beta) \rbrace
\end{eqnarray}

\par
From the first of the preceding equations we clearly recover the standard model result for $J_{\mu}(Z)$ in the limit $\omega^2\rightarrow 0$. The new neutral fermion's coupling with the standard Z is suppressed by a factor $\omega^2$. However, the standard fermion's couplings with the new Z' are not suppressed, according to the second of equations III.7. The main question at this point is how small $\omega$ can be in order that the model agrees with present data. It is well known that the presently known high precision data requires the inclusion of the standard model quantum corrections, in order to make a consistent comparison between theory and experiment. To compare the model with data we have considered that the new terms in equations III.7 are small corrections to the standard model predictions and at least of the same order of the quantum corrections. We have computed the corrections to the Z couplings to standard fermions and found

\bigskip
\begin{eqnarray}
g_V^{L,R} & = & g_V^{SM}-(\omega^2 \sin^4 \beta) \lbrace T^L_3 -2Q \rbrace \nonumber\\
g_A^{L,R} & = & g_A^{SM}-(\omega^2 \sin^4 \beta) \lbrace T^L_3 \rbrace \nonumber\\
\end{eqnarray}

\par
The Particle Data Group, in their 2000 edition \cite{RPP}, summarizes the present data from low energy lepton interaction, lepton-hadron collisions and the high precision data from LEP and SLAC. They also present the experimental averages for the $g_V$ and $g_A$ couplings for charged and neutral leptons. The most stringent bounds  comes from the 
effective coupling of the Z to the electron neutrino 
$g_{exp}^{\nu e} =0.528\pm0.085$ and
$\Gamma^{inv}_{exp}(Z)=498.8\pm 1.5$ MeV, to be compared with the
standard model predictions
 $g_{SM}=0.5042$ and $\Gamma^{inv}_{SM}(Z)=501.65\pm0.15$ MeV. For the
 muon neutrino coupling with the Z boson, the Particle Data Group
 quotes $g_{exp}^{\nu \mu} =0.502\pm 0.017$. 
 We have performed a fit to this data, using the standard model predictions, and we find that deviations from the standard model must be bounded, at $95\%$ confidence level by: 
\bigskip
\begin{equation}
(\omega^2 \sin^4 \beta ) < 10^{-4}
\end{equation}

\par
This bound is consistent with the present experimental constraint on the $\rho$ parameter.
With the value for $\sin\beta$ given in equation III.6, we have the bound
\begin{equation}
v_R > 30 \, v_L
\end{equation}

\par 
For the new Z' mass we have

\begin{equation}
M_{Z'} > 800 \quad GeV
\end{equation}

This value is a little above the present experimental bounds on new gauge bosons searches done by the CDF and D0 collaborations \cite{FAB} at Fermilab.

\section{The fermion masses}\setcounter{equation}{0}
 
\par

The fermion  mass spectrum depends both on the Higgs choice of the model and on the fundamental fermionic representation. A particular property of the model is the presence of left and right handed singlets in the fundamental representation. This means that we can add to the mass lagrangian new bare terms or new Higgs scalars which have no consequences on the vector gauge boson masses. We consider two new Higgs singlets, one that is coupled to Dirac terms in the mass lagrangian - $S_D$ - and the other that couples to Majorana terms - $S_M$. After spontaneous symmetry breaking they develop vacuum parameters $s_D$ and $s_M$, respectively.
\par
The most general Yukawa lagrangian is given by,
\begin{eqnarray}
{\cal L} & = & f \lbrace \overline{\ell_L} \tilde\chi_L \nu_R + \overline{L_R} \tilde \chi_R N_L \rbrace  + 
f' \lbrace \overline{\ell_L} \tilde\chi_L N_L^C + \overline{L_R} \tilde\chi_R \nu_R^C \rbrace \nonumber\\ &+&
gS_M \lbrace \overline{N_L^C} N_L + \overline{\nu_R^C}\nu_R \rbrace +g'S_D \lbrace \overline{\nu_R}N_L \rbrace \nonumber\\ &+& 
f" \lbrace \overline{\ell_L}\Phi L_R \rbrace + g^*S_D \lbrace \overline{e_R}E_L \rbrace + h.c. 
\end{eqnarray}

For the neutral fermions, the  mass lagrangian after symmetry breaking is given by
\begin{eqnarray}
{\cal L} & = & f v_L \overline{\nu_L} \nu_R + f'v_L\overline{\nu_L^c} N_L + fv_R\overline{N_R}N_L  + f'v_R \overline{N_R^C}\nu_R + \nonumber\\ &+&
gs_M\overline{N_L^C}N_L + gs_M\overline{\nu_R^C}\nu_R + \nonumber\\&+& g' s_D \overline{\nu_R}N_L + g" k \overline{\nu_L}N_R, 
\end{eqnarray}

\par
For the charged fermions we have a similar lagrangian, except for the conjugated terms. The generalization for the other families is straightforward. The diagonalization 
is most easily done \cite{TPC} by introducing the self conjugated fields 
\begin{eqnarray}
\chi_i & = &\psi_{iL}+\psi_{iL}^C \nonumber\\
\omega_j & = &\psi_{jR} +\psi_{jR}^C 
\end{eqnarray}
with $ i,j=\nu,N $.
\par
We will consider in this paper that the order of magnitude of the fermionic mass spectrum is given by the symmetry breaking scales and their combinations. This is a departure from the standard model procedure of adjusting the coupling constants in the Yukawa interactions. So we are supposing that all coupling constants are of order one.

In the basis $(\chi_\nu;\omega_N;\chi_N;\omega_\nu)$ the general neutrino mass matrix is:

$$ M_{\nu ,N}=\pmatrix{0&\displaystyle{k\over 2}&v_L&\displaystyle{v_L
\over 2}\cr \displaystyle{k\over 2}&0&\displaystyle{v_R\over 2}&v_R\cr  v_L&\displaystyle{v_R\over 2}&s_M&\displaystyle{{s_D}\over 2}\cr
\displaystyle{v_L\over 2}&v_R&\displaystyle{{s_D}\over 2}&s_M\cr}$$
and the charged fermion mass matrix is:
$$ M_{e,E}=\pmatrix{0&k'&0&v_L\cr
                    k'&0&v_R&0\cr
                    0&v_R&0&s_D\cr
                    v_L&0&s_D&0\cr}$$

\par
For this last case, we recover the Dirac formalism by the standard \cite{TPC} $\pi/4$ rotations over the Majorana fields.

\par
 
After the results of the Super Kamiokande collaboration \cite{SUK} there are strong experimental evidence that neutrinos oscillate and have non-zero mass \cite{RNM}. Atmospheric neutrino  experiments are consistent with $\nu_{\mu}\leftrightarrow\nu_{\tau}$ oscillations with a large mixing and $\Delta m^2=(1.5-6)10^{-3}$ $eV^2$. Solar neutrino evidence still allow some alternative solutions like vacuum oscillations and large and small angle MSW mechanism. If one considers the LSND results \cite{ATH} then the solution for the neutrino parameters seems to indicate the need of a forth sterile neutrino, but so far there is no confirmation of this result. In phenomenological models for quark and lepton masses we have a number of possibilities for fermion mass textures in models such as  $SO(10)$ \cite{MAT}.
In view of the present experimental situation we will not proceed to  fit  all the neutrino masses and mixings but we will look for solutions which could accommodate neutrinos with masses in the $10^{-2}-10^{-3}$ eV range. We present three possible solutions, which differ in the choice of the symmetry breaking parameters.
\par

\bigskip
$\bullet$ {\bf Model I}
\bigskip

\par
In this model we consider that the mixed Higgs field do not break the vacuum symmetry ($k=k'=0$) and that the Higgs singlets are both broken at $s_{GUT}$. For neutrinos we have the following masses
\begin{eqnarray}
m_{\nu 1}& = & v_L^2/s_{GUT}  \nonumber\\
m_{\nu 2}& = & v_R^2/s_{GUT}  \nonumber\\
m_{N1} & \simeq & m_{N2}\simeq s_{GUT}  \nonumber\\
\end{eqnarray}
and for the charged fermions
\begin{eqnarray}
m_e & = & v_L v_R/s_{GUT}  \nonumber\\
m_E & = & s_{GUT}
\end{eqnarray}
In this model we have the "universal see-saw" mass relation \cite{ADA} 
$m_{\nu1}m_{\nu2}=m_e^2$. Using  $ v_L=v_{Fermi}$ and $ s_{GUT}=10^{16}$ GeV, we must have 
$v_R=10^{10}$ in order to obtain the correct value for the electron mass. The first generation mass spectrum  is then given by $ m_{\nu1}\simeq10^{-2}$ eV ; $m_{\nu2}\simeq 10$ TeV ;  
$m_{N1}\simeq m_{N2}\simeq 10^{16}$ GeV; $ m_e \simeq 1$ MeV; $ M_E\simeq 10^{16}$ GeV. The smallness of the electron mass is a consequence of a "see-saw" mass relation given by equation IV.5. This is a departure for the standard model mechanism for fermion masses. 
There is no mixing between $\nu_1$ and $\nu_2$ and we have an example of a sterile neutrino coming from a new exotic doublet. The presently observed neutrino mixing must come from a possible generation mixing in the general neutrino mass lagrangian. Standard charged quarks are also found to be in the MeV mass range, with $m_q = v_L v_R/s_{GUT}$. With this high value for $v_R$ we have no experimental accessible accelerator possibilities for new gauge vector bosons. For the Large Hadron Collider (LHC) it has recently been shown \cite{YAC} that heavy neutrino production is limited to masses of a few hundreds of GeV. Heavy neutrinos with masses in the TeV region can be produced in the next generation \cite{GCV} of ${e^+}{e^-}$ or ${e^-}{\mu^+}$ colliders.
\par 
The $v_R$ value consistent with the electron mass is of the order of the Peccei-Quinn symmetry breaking scale \cite{JEK} and we can have a possible explanation of the small value for the $\theta$-angle of the strong CP problem, as shown in references \cite{SMB,KSB}. In order to generate the other families mass spectrum we have two possibilities. The first one, following the standard model procedure, is simply to adjust the coupling constants in the general mass lagrangians. The other possibility is to enlarge the Higgs singlet sector, postulating one new field for each family. In this last case we have an hierarchy for neutrino masses. 

\par

\vfill\eject
$\bullet$ {\bf Model II}
\bigskip

\par
In this model we impose explicit lepton number conservation and $ k=k'=0 $.The Majorana mass terms must be zero, except for the Higgs singlet coupled to Majorana mass terms in the neutrino sector  that can have two units of leptonic number. Lepton number is then spontaneously broken by this term at a scale $s_M \simeq s_{GUT}$ and the Higgs singlet coupled to Dirac mass terms is allowed to be broken at a lower scale. The neutrino mass spectrum is given by
\begin{eqnarray}
m_{\nu 1}& = & v_L^2/s_M  \nonumber\\
m_{\nu 2}& = & v_R^2/s_M  \nonumber\\
m_{N_1,N_2} & = & s_M\pm s_D/2  \nonumber\\
\end{eqnarray}
and for the charged fermions
\begin{eqnarray}
m_e & = & v_L v_R/s_D  \nonumber\\
m_E & = & s_D
\end{eqnarray}
If we take $v_L=v_{Fermi}$ and $s_M=s_{GUT}$, then the electron and quark masses allow a solution given by $v_R\simeq 10^4$ GeV and $s_D\simeq 10^{10}$ GeV. Here again we have a "see-saw" mechanism for the electron mass. In this model we could have an experimentally accessible new neutral current, as shown in Fig. 1. One light neutrino has a mass in the $10^{-2}-10^{-3}$ eV range and the other is in the $1 - 0.1$ eV region. These neutrinos are orthogonal and again we have a new sterile neutrino. The Peccei-Quinn symmetry breaking scale reappears in the Higgs singlet sector. Family replication can be recovered with different Dirac singlets and for only one Majorana mass scale we can have a degenerate neutrino mass spectrum.
\par
\bigskip
$\bullet$ {\bf Model III}
\bigskip

\par
We consider the same Higgs sector as in Model I, except for the mixed representation, with a small value for $k'$. The electron mass is changed to $ m_e = k'+ v_L v_R /s_{GUT}$. If we adjust $k'=m_e$ then there is no need to a large $v_R$ as in the first model. The down-quark mass is given by the same value of $k'$. However, for the up-quark mass we need a new $k\simeq k'$ in the mixed representation and this will not give a correct neutrino mass. An alternative to this difficulty is to consider that neutrinos are massless at tree level and acquire masses through quantum corrections \cite{KSB,GBA}. As this mechanism involves new additional hypothesis we will not develop the full solution here. We can recover the bound $v_R= 10^{3}-10^{4}$ GeV and we have the possibility to test this model through the new $Z'$ interactions.
\par
The generalization for the other families can be easily extended from the above arguments. However, the mixing angle pattern is not so simple. A first approach is to generalize the see-saw mechanism, with mixing angles given by the mass ratios $ \theta_{mix} \simeq m_{\nu}/m_N $, which are very small numbers. There are many models that avoid such a restriction: the introduction of an arbitrary number of right-handed neutrinos \cite{CEC}; some fine-tuning in the neutrino mass matrix \cite{BUC} and any general singular neutrino mass matrix can disconnect mixing parameters from mass ratios. So we will take mixing angles and neutrino masses as independent parameters. The high precision experimental data from LEP and SLC gives the bound \cite{NAR} $ \sin \theta_{mix}^2 < 10^{-2}-10^{-3}$.

\section{Phenomenological consequences}\setcounter{equation}{0}

\par
Let us now turn our attention to some of the phenomenological consequences of the models developed in the preceding section. For Model I the high value of $v_R$ makes new gauge bosons out of reach even for the next generation of accelerators. The model can be tested via the new mirror fermions, coupled to the standard model neutral current according to equation III.7. New charged leptons are coupled to the Z proportional to $ \sin^2 \theta_W $ and their phenomenology was extensively studied by many authors \cite{DJO}. The new neutral heavy lepton's coupling to Z is suppressed by a factor $\omega^2 \sin^4 \beta  < 10^{-4}$ and can  be detected only if heavy neutrinos are mixed with light neutrinos. This introduces a new mixing parameter that it is also known to be small \cite{NAR}, of the order of  $\alpha^2_{mix} \simeq 10^{-2}-10^{-3}$. The $N \bar N$ coupling to Z is suppressed by $\sin^2 \alpha_{mix}$ and the light-to-heavy neutrino coupling to Z is given by a single power of $\sin \alpha_{mix}$. This means that single heavy neutrino production is favored relative to pair production \cite{GCV,NOS}. 
\par
In Fig. 1 we show the total cross section for pair and single Dirac heavy neutrino production in $e^+e^-$ collisions at a new Next Linear Collider (NLC) at ${\sqrt s}=2$ TeV. For a single heavy neutrino production we can identify the signal by it's decay in ${e^\pm}{W^\mp}$. These processes can be readily calculated by using high energy algebraic programs such as CompHep \cite{HEP}. 
\par
In Fig. 2 we show the $e W$ invariant mass distribution for both the signal and standard model background for $ e^+ e^- \rightarrow e^{\pm}\, \nu_e \, W^{\mp} $ and $M_{Ne}=$ 400 GeV. In this process we have 6 diagrams from the new heavy lepton and 12 from the standard model background \cite{HEP}. Here again we can separate the signal from the standard model background, if we take the upper bound $\sin^2 \alpha_{mix} < 0.0052$ \cite{YAC}.
\par
In Model II we can have the production of new heavy neutral gauge bosons with mass scales accessible at hadron colliders. Dilepton production in hadron-hadron collisions give a very clear signal for a new Z'. The new facilities at the Tevatron proton-antiproton collider, with $\sqrt s=2$ TeV and a luminosity of 1000 pb$^{-1}$, will allow the search of the new Z' with masses in the 800-1000 GeV region. For $p \bar p \longrightarrow Z' \longrightarrow \ell^{\pm}\ell^{\mp}$, with $\ell= e,\mu$, we expect 20 events for a Z' mass of 800 GeV. We have employed the parton distribution functions CTEQ4m by \cite{CTQ}.
\par
The Large Hadron Collider (LHC) facilities using proton-proton collisions at $\sqrt s= 14$ TeV, will attain higher Z' masses in the 1-4 TeV region. We show in Fig. 3 the invariant mass distribution for the $Z'$ decay at the LHC, with $M_{Z'}= 2$ TeV in the leptonic channel $Z'\rightarrow \ell^{\pm}\ell^{\mp}$ with $ \ell=e,\mu $. For an integrated luminosity at LHC of 100 fb$^{-1}$ and cuts of $E_{\ell}> 20$ GeV and ${|\eta|}< 2,5$ we expect 1000 events for $M_{Z'}=1$ TeV and only one event for $M_{Z'}=4$ TeV. 
New heavy neutrinos can also be coupled to new heavy neutral gauge bosons and can be produced at the new large hadron collider. A detailed study of these possibilities was done in Ref.  \cite{ATL}.
\par

\section{Conclusions}
\par
In conclusion we have shown that spontaneously broken parity models with a consistent fermion mass spectrum can be built at intermediate scales ranging from $10^3$ to $10^{10}$ GeV. New mirror fermions are present and can be connected to ordinary fermions through neutral currents. We propose a simple  Higgs sector for the model that allows several physically interesting solutions. The charged fermion masses can be generated by a "see-saw" mass relation, analogous to the neutrino sector. Various scenarios for neutrino masses are possible and a more clear experimental definition on neutrino masses will allow to test their reality. Neutral heavy gauge bosons Z and Z', coupled to ordinary and new mirror fermions, are expected to play a fundamental role in the understanding of the left-right symmetry. This is a departure from other models \cite{ZGB,KOB} that propose gravitation as the connection between the left and right sectors. Experimental consequences of the model could be found at the next generation 
of colliders. 
\vskip 1.0cm
\par
{\it Acknowledgments:} This work was partially supported by the
following Brazilian agencies: CNPq, FUJB, FAPERJ and FINEP.

\vspace{0.5cm}
\LARGE
{\bf Figure Captions}
\normalsize
\begin{enumerate}
\item Total cross section for pair and single heavy neutrino production for ${e^+}{e^-}$ collisions at NLC collider with ${\sqrt s}=2$ TeV. $N_e$ corresponds to a heavy neutrino of the electron family.
\item Invariant $e W$ mass distribution in the process ${e^+}{e^-}\rightarrow {e^+} \,{{\nu}_e} \,{W^-} $ at NLC with ${\sqrt s}=2$ TeV and $M_{Ne}= 400$ GeV. The flat part of the curve is the standard model background.
\item The invariant mass distribution for $p p \longrightarrow Z'\rightarrow \ell^{\pm}\ell^{\mp}$ (where $\ell=$e or $\mu$) for pp collisions at LHC with ${\sqrt s}=14$ TeV, $E >20$ GeV and ${|\eta|}< 2,5$ and $M_{Z'}=2$ TeV.

\end{enumerate}
\vfill\eject
\end{multicols}

\begin{thebibliography}{ABC}

\bibitem{BBR} B.Brahmachari and R.N.Mohapatra, hep-ph/9805429 v2.
\bibitem{ZGB} Z.G.Berezhiani and R.N.Mohapatra, Phys. Rev. D {\bf 52} (1995) 6607.
\bibitem{SMB} S.M.Barr, D.Chang and G.Senjanovi\u c, Phys. Rev. Lett. {\bf 67} (1991) 2765.
\bibitem{RFO} R.Foot and R.R.Volkas, Phys. Rev. D {\bf 52} (1995) 6595.
\bibitem{JCP} J.C.Pati and A.Salam, Phys. Rev. D {\bf 10} (1974) 275; R.N.Mohapatra and J.C.Pati,Phys. Rev. D {\bf 11} (1975) 566; G.Senjanovi\u c and R.N.Mohapatra, Phys. Rev. D {\bf            12} (1975) 1502; R. N. Mohapatra and R.E. Marshak, Phys. Lett. B {\bf91} (1980) 222. An             extensive list of references can be found in  R.N. Mohapatra and P.B. Pal, "Massive             Neutrinos in Physics and Astrophysics", World Scientific, Singapore, 1998.
\bibitem{SLG} S.L.Glashow, Phys. Lett. B {\bf 167} (1986) 35; E.D.Carlson and S.L.Glashow, Phys.Lett. B {\bf 193} (1987) 168.
\bibitem{SEN} G. Senjanovi\u c, Nuc. Phys. B {\bf153} (1979) 334.
\bibitem{RPP} Review of Particle Physics, Eur. Phys. J. C {\bf15} (2000) 1.
\bibitem{FAB} F.Abe et al, Phys. Rev. Lett. {\bf 79} (1997) 2192; S.Obachi et al, Phys. Lett.                            B {\bf 385} (1996) 471.
\bibitem {TPC} T.P.Cheng and L.F.Li, Gauge Theory of Elementary Particle Physics. Oxford, 1984.
\bibitem {SUK} A. Mann, hep-ex/9912007.
\bibitem {RNM} R.N.Mohapatra, hep-ph/9910365; K.Zuber, Phys. Rept. C {\bf305} (1998) 295.
\bibitem {ATH} C. Athanassopoulos et al., Phys. Rev. Lett. {\bf 77} (1996) 3082.
\bibitem {MAT} K.Matsuda, T.Fukuyama and H.Nishiura, Phys. Rev. D {\bf 61} (2000) 053001; H.  Fritsch, Nuc. Phys. B {\bf 155} (1979) 189; P.Ramond, R.G.Roberts and G.G.Ross, Nuc.                                                   Phys. B {\bf 406}(1993) 19; G.C.Branco, L.Lavoura and F.Mota, Phys. Rev. D {\bf 39}(1989)          3443.
\bibitem {ADA} A.Davison and K.C.Wali, Phys. Rev. Let. {\bf 59} (1987) 393.
\bibitem {YAC} Y.A.Coutinho, J.A.Martins Sim\~oes, P.P.Queiroz Filho, Phys. Rev D {\bf 57} (1998) 6975; F.M.L.Almeida Jr., Y.A.Coutinho, J.A.Martins Sim\~oes, M.A.B. Vale, Phys. Rev. D {\bf 62} (2000) 075004 (6 pages). 

\bibitem {GCV} G.Cveti\u c and C.S.Kim, Phys. Lett. B {\bf 461} (1999) 248 and Phys. Lett.
		B {\bf 471} (2000) 471; F.M.L.Almeida Jr., Y.A.Coutinho, J.A.Martins Sim\~oes, M.A.B.                         Vale, hep-ph 0008231
\bibitem {JEK} J.E.Kim, Phys. Rep. {\bf 150} (1987) 1.
\bibitem {KSB} K.S.Babu and R.N.Mohapatra, Phys. Rev. D {\bf 41} (1990) 1286.
\bibitem {GBA} G.Barenboim and F.Scheck, Phys. Lett. B {\bf 461} (1999) 235.
\bibitem {CEC} C.Jarlskog, Phys. Lett. B {\bf241} (1990) 579.
\bibitem {BUC} W.Buchm\"uller and D.Wyler, Phys. Lett. B {\bf249} (1990) 458; W.Buchm\"uller and C.Greub, Nuc. Phys. B {\bf363} (1991) 345.
\bibitem {NAR} E.Nardi, E.Roulet and D.Tommasini, Phys.Lett. B{\bf327} (1994) 319;
          P.Bamert, C.P.Burgess and I.Maksymyk, Phys. Lett. B{\bf267} (1995) 282.
\bibitem{DJO} A.Djouadi,J.Ng and T.G.Rizzo in: Electroweak Symmetry Breaking
          and Beyond the Standard Model. Ed.T.Barklow, S.Dawson, H.E.Haber 
          and S.Siegrist, World Scientific, Singapore. hep-ph 9504210;\par
          P.Langacker,M.Luo and A.K.Mann, Rev. Mod. Phys. {\bf64} (1992) 87.
\bibitem {NOS} F.M.L. Almeida Jr., J.H.Lopes, J.A.Martins Sim\~oes and C.M.Porto, Phys. Rev.          D{\bf44}  (1991) 2836 ; F.M.L.Almeida Jr., J.H.Lopes, J.A.Martins Sim\~oes, P.P.Queiroz                        Filho and A.J.Ramalho Phys. Rev. D{\bf51} (1995) 5990.
\bibitem {HEP} A.Pukhov, E.Boss, M.Dubinin, V.Edneral, V.Ilyin, D.Kovalenko, A.Krykov, V.Savrin,          S.Shichanin and A.Semenov, "CompHEP"- a package for evaluation of Feynman diagrams and          integration over multi-particle phase space. Preprint INP MSU 98-41/542, hep-ph/9908288.
\bibitem {CTQ} H.L.Lai, J.Huston, S.Kuhlmann, F.Olness, J.Owens, D.Soper, W.K.Tung, H.Weerts, Phys. Rev. D{\bf55} (1997) 1280.
\bibitem {ATL} A.Ferrari, J.Collot, M-L.Andrieux, B.Belhorma, P.de Saintignon, J-Y.Hostachy, Ph. Martin and M.Wielers, Phys. Rev. D{\bf62} (2000) 013001.
\bibitem {KOB} I.Kobzarev, L.Okun and I.Ya Pomeranchuk, Sov. Jour. Nuc. Phys. {\bf3} (1966) 837.
 
\end{thebibliography}
\end{document}